\documentclass[lettersize,journal]{IEEEtran}
\usepackage{amsmath,amsfonts}
\usepackage{algorithmic}
\usepackage{algorithm}
\usepackage{array}
\usepackage[caption=false,font=normalsize,labelfont=sf,textfont=sf]{subfig}
\usepackage{textcomp}
\usepackage{stfloats}
\usepackage{url}
\usepackage{verbatim}
\usepackage{graphicx}
\usepackage{cite}
\usepackage{bm}
\usepackage{amsmath}
\usepackage{amssymb}
\usepackage{amsfonts}
\usepackage{lipsum}
\begin{document}

\title{SNR Maximization and Localization for UAV-IRS-Assisted Near-Field Systems}
\author{Hanfu Zhang, Yidan Mei, Erwu Liu, \emph{Senior Member, IEEE}, and Rui Wang, \emph{Senior Member, IEEE}
\thanks{
This work has been submitted to the IEEE for possible publication. Copyright may be transferred without notice, after which this version may no longer be accessible.

The authors are with the College of Electronic and Information Engineering, Tongji University, Shanghai 201804, China (e-mail: hanfuzhang@tongji.edu.cn; 2130690@tongji.edu.cn; erwu.liu@ieee.org; ruiwang@tongji.edu.cn).
}}

\maketitle

\begin{abstract}
This letter introduces a novel \emph{unmanned aerial vehicle} (UAV)-\emph{intelligent reflecting surface} (IRS) structure into near-field localization systems to enhance the design flexibility of IRS, thereby obtaining additional performance gains. Specifically, a UAV-IRS is utilized to improve the harsh wireless environment and provide localization possibilities. To improve the localization accuracy, a joint optimization problem considering UAV position and UAV-IRS passive beamforming is formulated to maximize the receiving \emph{signal-to-noise ratio} (SNR). An alternative optimization algorithm is proposed to solve the complex non-convex problem leveraging the \emph{projected gradient ascent} (PGA) algorithm and the principle of minimizing the phase difference of the receiving signals. Closed-form expressions for UAV-IRS phase shift are derived to reduce the algorithm complexity. In the simulations, the proposed algorithm is compared with three different schemes and outperforms the others in both receiving SNR and localization accuracy.
\end{abstract}


\section{Introduction}
\IEEEPARstart{W}{ith} the development of the \emph{fifth generation mobile networks} (5G) and its empowering applications across various industry progresses, emerging services such as mobile \emph{Internet of things} (IoT) and  \emph{artificial intelligence} (AI) continually arise. As a result, there is an increasing demand for centimeter-level localization accuracy on the \emph{sixth generation mobile networks} (6G) \cite{bourdoux20206g, dang2020should, letaief2019roadmap}. Although traditional \emph{global navigation satellite systems} (GNSSs) have provided localization services with broad coverage and massive connectivity \cite{kaplan2017understanding}, their precision falls short of future requirements. In addition, they fail to deliver stable, high-quality services in environments like urban canyons. Recently, wireless communication signals-assisted localization has surged to tackle such GNSS-denied issues and achieve higher localization accuracy. For instance, Shahmansoori \emph{et al.} \cite{shahmansoori2017position} proposed a two-stage algorithm based on multiple measurement vectors matching pursuit and space-alternating generalized expectation maximization algorithm. Nazari \emph{et al.} \cite{nazari2023mmwave} used downlink \emph{multiple-input multiple-output} (MIMO) \emph{orthogonal frequency-division multiplexing} (OFDM) signals to estimate an unsynchronized multi-antenna \emph{user equipment's} (UE's) \emph{three-dimensional} (3D) position and 3D orientation. However, signals can still be obstructed by physical barriers in certain circumstances, i.e., the \emph{line-of-sight} (LoS) path is blocked. Under this condition, achieving high-precision localization becomes nearly impossible.

To tackle this issue, the \emph{intelligent reflecting surface} (IRS) \cite{wu2021intelligent, yuan2021reconfigurable, wu2019towards} has been integrated into communication-aided localization systems. An IRS is a passive reflecting device composed of multiple reflecting elements. Localization can be realized with the help of \emph{virtual LoS} (VLoS) paths provided by IRS even when the LoS path is blocked and can be improved via the passive beamforming process. In this process, the phase shift for each reflecting element is designed, and the wireless channels can be manipulated. For example, Dardari \emph{et al.} \cite{dardari2021nlos} proposed two practical signaling and localization algorithms based on OFDM downlink systems, as well as methods to design the IRS time-varying reflection coefficients. Zhang \emph{et al.} \cite{zhang2023near} developed a joint estimation algorithm considering UE position and realistic impairments in OFDM systems and proposed a \emph{semidefinite relaxation} (SDR)-based IRS phase shift optimization algorithm to enhance localization performance.

To realize greater design flexibility of IRS for the further enhancement of wireless environments, a novel \emph{unmanned aerial vehicle} (UAV)-IRS architecture has recently gained attention. This architecture features a UAV equipped with an IRS. The UAV position and UAV-IRS phase shift can both be controlled by the \emph{base station} (BS) through a signaling channel. For instance, Wei \emph{et al.} \cite{wei2023average} analyzed the average error probability in UAV-IRS-assisted short packet communication systems. Wang \emph{et al.} \cite{wang2022covert} proposed a UAV-IRS-aided covert communication scheme to maximize the covert communication rate. However, current research about UAV-IRS mainly focuses on its enhancement to communication systems, and to the best of the authors' knowledge, there's no research considering UAV-IRS-assisted localization systems. To fill this gap, the main contributions of this letter are summarized below:

\begin{itemize}

    \item We introduce a novel UAV-IRS-assisted near-field localization model, where \emph{maximum likelihood estimation} (MLE) algorithm is adopted to realize high-precision position estimation. To further improve the localization accuracy, an optimization problem for UAV position and UAV-IRS phase shift is formulated, which maximizes the receiving \emph{signal-to-noise ratio} (SNR).
    
    \item We design an alternative algorithm to solve the non-convex SNR maximization problem. A \emph{projected gradient ascent}(PGA)-based method is used to update the UAV position. Closed-form expressions for UAV-IRS phase shift are derived to minimize the phase difference of the received signals, significantly reducing the algorithm complexity.

    \item We conduct simulations comparing our proposed optimization algorithm with three other design schemes. Extensive results demonstrate the effectiveness of our algorithm and the performance improvement in near-field localization.
\end{itemize}

\emph{Notations}: $p(\cdot)$ denotes the \emph{probability density function} (PDF). $\mathbf{M}^\top$ and $\mathbf{M}^*$ denote the transpose and conjugate of matrix $\mathbf{M}$, respectively. $\bm{v}[n]$ denotes the $n$th entry of vector $\bm{v}$, while $\bm{v}^{[q]}$ denotes the value of $\bm{v}$ in the $q$th iteration. $\mathbf{I}_M$ denotes an $M \times M$ sized identity matrix. $\mathbb{C}^{m \times n}$ denotes $m \times n$ matrices with complex entries. $\Vert \cdot \Vert$, $| \cdot |$ and ${\rm arg}(\cdot)$ represent calculating 2-norm, modulus and argument, respectively. Finally, $\mathcal{CN} (\bm{m}, \mathbf{C})$ represents complex Gaussian distributions with a mean vector of $\bm{m}$ and a covariance matrix of $\mathbf{C}$.

\section{System Model and Problem Formulation}

\begin{figure}[!t]
\centering
\includegraphics[width=0.9 \linewidth]{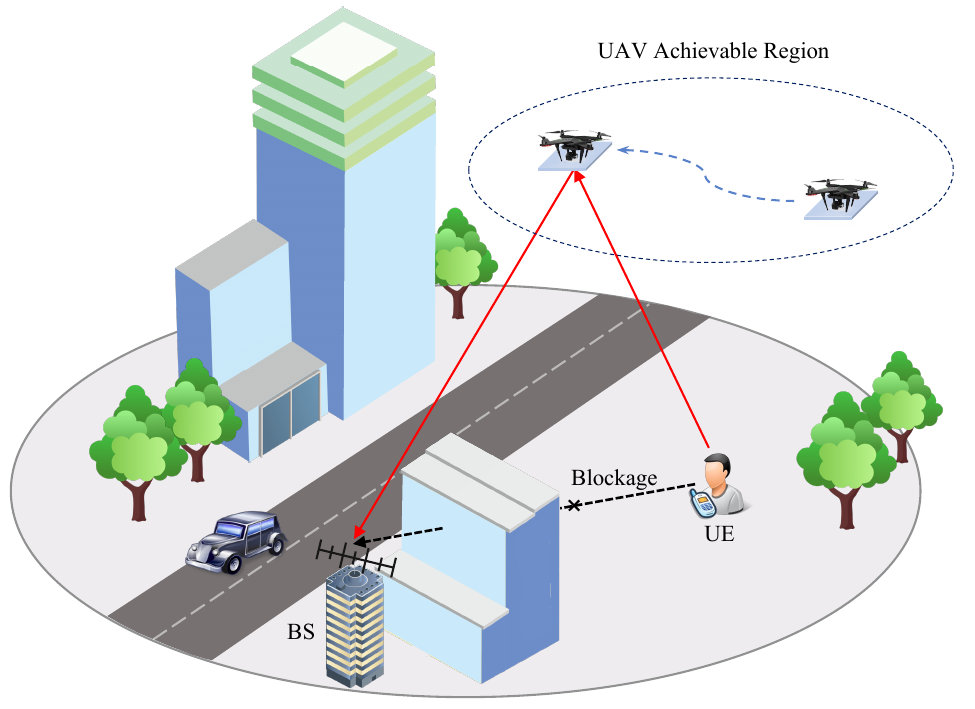}
\caption{UAV-IRS-assisted near-field localization system, where the LoS path is blocked.}
\label{FigSystemModel}
\end{figure}

\subsection{Signal Model}

As illustrated in Fig. \ref{FigSystemModel}, the considered UAV-IRS-assisted localization system is composed of a single-antenna UE, a UAV-IRS equipped with $N_I$ reflecting elements, and a BS equipped with $N_B$ receiving antennas. The LoS path between UE and BS is blocked, thus only the VLoS path provided by UAV-IRS can be utilized to carry out localization. Both UAV-IRS elements and BS antennas are arranged in \emph{uniform linear arrays} (ULAs) with element/antenna spacing of $d = \lambda / 2$, where $\lambda$ denotes the wavelength. Although ULAs are considered in this letter, it's worth noting that the proposed algorithms can be directly extended to \emph{uniform planar arrays} (UPAs). The UE is assumed to be static or moving slowly, and the UAV-IRS can move in a certain region to further improve the path quality and localization accuracy.

In our work, the positions of the BS and UAV-IRS are assumed to be known, whose reference point coordinates can be denoted as $\bm{p}_B$ and $\bm{p}_I$, respectively, and the coordinate of their antennas/elements can be expressed by $\bm{p}_b$ and $\bm{p}_i$, respectively, where $b = 1, 2, \cdots, N_B$ and $i = 1, 2, \cdots, N_I$. The position of UE $\bm{p}_u$ is assumed to be unknown and is the parameter to be estimated.

One single symbol $s$ is utilized in the localization process in this paper and the received signal $\bm{y} \in \mathbb{C}^{N_B \times 1}$ at BS can be expressed as:
\begin{equation}
    \bm{y} = \sqrt{P} \bm{h} s + \bm{n},
\end{equation}
where $P$ is the transmit power, $\bm{n}$ is the additive Gaussian noise at BS, and $\bm{h}$ is the UE-IRS-BS channel, modeled as:
\begin{equation}
    \bm{h} = \sqrt{\frac{\kappa}{1 + \kappa}} \Bar{\bm{h}} + \sqrt{\frac{1}{1 + \kappa}} \Tilde{\bm{h}},
\end{equation}
where $\kappa$ is the Rician factor, $\Bar{\bm{h}}$ and $\Tilde{\bm{h}}$ are LoS and \emph{Non LoS} (NLoS) parts of the channel, respectively. Each term of $\Bar{\bm{h}}$ is given by:
\begin{equation}
    \label{LoSChannel}
    \Bar{\bm{h}}[b] = \sqrt{\rho_{u,I} \rho_{I,B}} \sum_{i=1}^{N_I} \bm{w}[i] e^{-j \frac{2 \pi}{\lambda} (d_{b,i} + d_{i,u}) },
\end{equation}
where $\sqrt{\rho_{u, I}} = \frac{\sqrt{G_u} \lambda}{4 \pi \Vert \bm{p}_I - \bm{p}_u \Vert}$ and $\sqrt{\rho_{I, B}} = \frac{\sqrt{G_B} \lambda}{4 \pi \Vert \bm{p}_B - \bm{p}_I \Vert}$ are the free-space path losses between UE and UAV-IRS, and between UAV-IRS and BS, respectively, with $G_u$ and $G_B$ being the antenna gains at UE and BS. $\bm{w}[i]$ is the phase shift of the $i$th element of the UAV-IRS. $d_{b,i} = \Vert \bm{p}_i - \bm{p}_b \Vert$ represents the distance between the $i$th element of UAV-IRS and the $b$th antenna of BS, while $d_{i,u} = \Vert \bm{p}_u - \bm{p}_i \Vert$  represents the distance between UE and the $i$th element of UAV-IRS.

Combining the Gaussian distributed NLoS portion of $\bm{h}$ with noise $\bm{n}$, the received signal can be rewritten as:
\begin{equation}
    \label{EqSignalModel}
    \bm{y} = \sqrt{P} \Bar{\bm{h}} s + \Tilde{\bm{n}},
\end{equation}
where the combined noise $\Tilde{\bm{n}}$ can be represented as:
\begin{equation}
    \Tilde{\bm{n}} \triangleq \sqrt{\frac{P}{1 + \kappa}} \Tilde{\bm{h}} s + \bm{n},
\end{equation}
and follows the distribution $\Tilde{\bm{n}} \sim \mathcal{CN}(\bm{0}, \sigma^2 \mathbf{I}_{N_B})$.

\subsection{UAV-IRS-Assisted Localization Problem}
According to (\ref{EqSignalModel}), the received signal $\bm{y}$ can be statistically characterized as:
\begin{equation}
    \bm{y} \sim \mathcal{CN}\left(\sqrt{\frac{\kappa P}{1 + \kappa}} \Bar{\bm{h}}, \sigma^2 \mathbf{I}_{N_B}\right),
\end{equation}
hence, according to the MLE criteria, the UE position $\bm{p}_u$ can be estimated by minimizing the following negative log-likelihood function:
\begin{align}
    L(\bm{p}_u) &\propto - \log p(\bm{y} \mid \bm{p}_u) \notag \\
    &= \frac{1}{\sigma^2} \Vert \bm{y} - \sqrt{\frac{\kappa P}{1 + \kappa}} \Bar{\bm{h}} \Vert^2.
\end{align}

Then, the estimate of UE position $\hat{\bm{p}}_u$ can be given by:
\begin{equation}
    \label{MLE}
    \hat{\bm{p}}_u = \arg \min_{\bm{p}_u}  L(\bm{p}_u).
\end{equation}

\emph{Remark 1:} According to (\ref{LoSChannel}), the parameters to be estimated appear in the $e$ exponent, and there is a large coefficient $2 \pi / \lambda$ before the parameters because of the high carrier frequency. This leads to severe oscillations of the likelihood function with respect to the position parameters, resulting in numerous local minima. Therefore, all gradient-based methods, such as gradient descent and Newton's method, are inapplicable here. We employ the simplest grid search method to solve (\ref{MLE}) thereby providing an estimate of the position parameters.

\subsection{SNR Maximization Problem}
To improve the localization accuracy, we formulate an optimization problem for the UAV position $\bm{p}_I$ and UAV-IRS phase shift $\bm{w}$. Receiving SNR is considered the objective function here, which is commonly used in existing localization works \cite{wang2023joint, liu2022joint}. According to the signal model (\ref{EqSignalModel}), the receiving SNR at BS can be expressed by:
\begin{equation}
    {\rm SNR}(\bm{p}_I, \bm{w}) = \frac{\kappa P}{(1 + \kappa) \sigma^2} \Vert \Bar{\bm{h}} \Vert^2,
\end{equation}
which depends on both $\bm{p}_I$ and $\bm{w}$. The SNR maximization problem can be formulated as:
\begin{align}
    \max_{\bm{p}_I, \bm{w}} & \quad {\rm SNR}(\bm{p}_{I}, \bm{w})  \label{originOptimizationProblem} \\
    s.t. \quad &| \bm{w}[i] | = 1,  i=1, \cdots, N_I,  \label{OOPa}\tag{\ref{originOptimizationProblem}a} \\
    & \Vert \bm{p}_I - \bm{p}_0 \Vert \leq R_{max}, \label{OOPb}\tag{\ref{originOptimizationProblem}b}
\end{align}
where (\ref{OOPa}) is the unit modulus constraint for UAV-IRS, and constraint (\ref{OOPb}) is to guarantee the UAV can only move in a circular achievable region centered at $\bm{p}_0$ with radius $R_{max}$. However, problem (\ref{originOptimizationProblem}) is difficult to solve due to the non-convex objective function as well as constraint (\ref{OOPa}) and coupled variables. The proposed solution to this problem will be given in the next section.

Integrating the two problems above, the complete localization process contains three steps. Firstly, a rough estimation of the UE position is obtained through MLE using the initial UAV position and random UAV-IRS phase shift. Secondly, the SNR maximization problem is solved based on the estimation result, and a better UAV position and UAV-IRS phase shift can be obtained. Both parameters are transmitted to the UAV via a reliable signaling channel. After the UAV reaches the designed position and configures the phase shift, UE transmits another sensing signal. Thirdly, the MLE is executed again at BS to obtain a more accurate UE position estimate.

\section{Proposed SNR Maximization Algorithm}
This section is devoted to solving the problem (\ref{originOptimizationProblem}) by developing a novel alternative optimization approach to optimize UAV position $\bm{p}_I$ and UAV-IRS phase shift $\bm{w}$. First, with fixed $\bm{w}$, we optimize $\bm{p}_I$ by proposing a PGA-based algorithm. Then, with fixed $\bm{p}_I$, we optimize $\bm{w}$ by converting the origin non-convex problem to a phase difference minimization problem. We derive closed-form expressions for the optimal phase shift, thus decreasing the algorithm complexity significantly. Finally, we provide the overall optimization algorithm.

\subsection{Optimization of $\bm{p}_I$ with Fixed $\bm{w}$}
\label{optimizePI}
Considering initially given or calculated $\bm{w}$ in the past iteration round, we optimize the position of UAV $\bm{p}_I$ by solving the following subproblem:
\begin{align}
\label{subproblem-pi}
    & \max_{\bm{p}_I} \quad {\rm SNR}(\bm{p}_{I}) \notag \\
    & s.t. \quad \Vert \bm{p}_I - \bm{p}_0 \Vert \leq R_{max},
\end{align}
which is non-convex because of the complex expression of SNR. We utilize the PGA algorithm to iteratively solve a better UAV position and obtain a higher SNR. In the $p$th iteration round, $\bm{p}_I^{[p]}$ can be obtained by calculating:
\begin{equation}
\label{updateP}
    \bm{p}_I^{[p]} = \mathcal{P} \left\{ \bm{p}_I^{[p - 1]} + \frac{\partial {\rm SNR}(\bm{p}_I^{[p-1]})}{\partial \bm{p}_I} k_p \right\},
\end{equation}
where $k_p$ represents the update step, and $\mathcal{P}\left\{ \cdot \right\}$ represents the projection operator, which can project parameters exceeding the feasible zone back into it. For any given position $\bm{p}$, $\Tilde{\bm{p}} \triangleq \mathcal{P} \left\{ \bm{p} \right\}$ can be obtained via solving:
\begin{align}
\label{projection}
    & \min_{\Tilde{\bm{p}}} \quad \Vert \Tilde{\bm{p}} - \bm{p} \Vert \notag \\
    & s.t. \quad \Vert \Tilde{\bm{p}} - \bm{p}_0 \Vert \leq R_{max},
\end{align}
which is convex, and can be directly solved using convex optimization solvers, such as CVX \cite{grant2014cvx}.

Take $\bm{p}_I[1]$ for example, $\frac{\partial {\rm SNR}(\bm{p}_I)}{\partial \bm{p}_I[1]}$ can be calculated as follows:
\begin{align}
    \frac{\partial {\rm SNR}(\bm{p}_I)}{\partial \bm{p}_I[1]} = \frac{\kappa P}{(1 + \kappa) \sigma^2} \sum_{b=1}^{N_B} \left( \frac{\partial \Bar{\bm{h}}[b]}{\partial \bm{p}_I[1]} \Bar{\bm{h}}^*[b] + \Bar{\bm{h}}[b] \frac{\partial \Bar{\bm{h}}^*[b]}{\partial \bm{p}_I[1]} \right),
\end{align}
where
\begin{equation}
    \frac{\partial \Bar{\bm{h}}[b]}{ \partial \bm{p}_I[1] } = \sqrt{\rho_{u,I} \rho_{I,B}} \sum_{i=1}^{N_I} \alpha_{i, u} \bm{w}[i] e^{-j \frac{2 \pi}{\lambda} (d_{b,i} + d_{i,u}) },
\end{equation}
and $\frac{\partial \Bar{\bm{h}}^*[b]}{ \partial \bm{p}_I[1] }  = \left( \frac{\partial \Bar{\bm{h}}[b]}{ \partial \bm{p}_I[1] } \right)^*$, with
\begin{align}
    \alpha_{i, u} = &- \left( \frac{\bm{p}_I[1] - \bm{p}_B[1]}{\Vert \bm{p}_I - \bm{p}_B \Vert^2} + \frac{\bm{p}_I[1] - \bm{p}_u[1]}{\Vert \bm{p}_u - \bm{p}_I \Vert^2} \right) \notag \\
    &- j \frac{2 \pi}{\lambda} \left( \frac{\bm{p}_i[1] - \bm{p}_b[1]}{\Vert \bm{p}_i - \bm{p}_b \Vert^2} + \frac{\bm{p}_i[1] - \bm{p}_u[1]}{\Vert \bm{p}_u - \bm{p}_I \Vert^2} \right).
\end{align}

The partial derivatives of ${\rm SNR}(\bm{p}_I)$ with respect to $\bm{p}_I[2]$ and $\bm{p}_I[3]$ can be similarly derived, and will not be further elaborated here.

\subsection{Optimization of $\bm{w}$ with Fixed $\bm{p}_I$}
\label{optimizeW}
Then, considering given $\bm{p}_I$, we optimize the phase shift of UAV-IRS $\bm{w}$ by solving the following subproblem: 
\begin{align}
\label{subproblem-pi}
    & \max_{\bm{w}} \quad {\rm SNR}(\bm{w}) \notag \\
    & s.t. \quad | \bm{w}[i] | = 1,  i=1, \cdots, N_I,
\end{align}
which is also non-convex due to the unit-module constraint. To tackle this problem, inspired by \cite{elzanaty2021reconfigurable}, by defining $\Tilde{\bm{w}} \triangleq \arg (\bm{w})$, we transfer (\ref{subproblem-pi}) to minimizing the sum of the square distance of the phases from their related centroid $\phi(\Tilde{\bm{w}})$ by:
\begin{equation}
    \min_{\Tilde{\bm{w}}} \gamma(\Tilde{\bm{w}}) \triangleq \sum_{b=1}^{N_B} \sum_{i=1}^{N_I} \left( \Tilde{\bm{w}}[i] - \frac{2 \pi}{\lambda} (d_{b, i} + d_{i, u}) - \phi(\Tilde{\bm{w}}) \right)^2,
\end{equation}
which is a convex function, and $\phi(\Tilde{\bm{w}})$ is given by:
\begin{equation}
    \phi(\Tilde{\bm{w}}) = \frac{1}{N_I N_B} \sum_{b=1}^{N_B}\sum_{i=1}^{N_I} \left( \Tilde{\bm{w}}[i] - \frac{2 \pi}{\lambda} (d_{b, i} + d_{i, u}) \right).
\end{equation}

For certain $k = 1, 2, \cdots, N_I$, $\gamma_k(\Tilde{\bm{w}})$ can be re-expressed by:
\begin{align}
    \gamma_k(\Tilde{\bm{w}}) =& \sum_{b=1}^{N_B} \sum_{i=1, i \neq k}^{N_I} \left( \Tilde{\bm{w}}[i] - \frac{2 \pi}{\lambda} (d_{b, i} + d_{i, u}) - \phi(\Tilde{\bm{w}}) \right)^2 \notag \\
    & + \sum_{b=1}^{N_B} \left( \Tilde{\bm{w}}[k] - \frac{2 \pi}{\lambda} (d_{b, k} + d_{k, u}) - \phi(\Tilde{\bm{w}}) \right)^2,
\end{align}
and by operating:
\begin{align}
    &\frac{\partial \gamma(\Tilde{\bm{w}})}{\partial \Tilde{\bm{w}}[k]} = 2 \left(1 - \frac{1}{N_I}\right) \sum_{b=1}^{N_B} \left( \Tilde{\bm{w}}[k] - \frac{2 \pi}{\lambda} (d_{b, k} + d_{k, u}) - \phi(\Tilde{\bm{w}}) \right) \notag \\
    &-2 \sum_{b=1}^{N_B} \sum_{i=1, i \neq k}^{N_I} \frac{1}{N_I} \left( \Tilde{\bm{w}}[i] - \frac{2 \pi}{\lambda} (d_{b, i} + d_{i, u}) - \phi(\Tilde{\bm{w}}) \right) = 0,
\end{align}
the optimal solution $\Tilde{\bm{w}}^\star[k]$ can be solved in a closed form:
\begin{equation}
\label{updateW}
    \Tilde{\bm{w}}^\star[k] = \frac{2 \pi}{\lambda N_B} \sum_{b=1}^{N_B} \left( d_{b,k} + d_{k,u} - \frac{1}{N_I} \sum_{i=1}^{N_I} (d_{b,k} + d_{k,u}) \right).
\end{equation}

Thus, the optimal UAV-IRS phase shift can be represented as:
\begin{equation}
    \bm{w}^\star = \left[ \exp (j \Tilde{\bm{w}}^\star[1]), \exp (j \Tilde{\bm{w}}^\star[2]), \cdots, \exp (j \Tilde{\bm{w}}^\star[N_I]) \right]^\top.
\end{equation}

In this way, signals reflected by different elements on UAV-IRS can arrive at BS with the minimum phase difference, thereby achieving higher receiving SNR.

\subsection{Overall Algorithm}

The above analysis of SNR assumes the known UE's position. In practice, the estimate of UE position $\hat{\bm{p}}_u$ is used instead. Thus, we define a new function $\Gamma(\bm{p}_I, \bm{w}, \hat{\bm{p}}_u)$ to replace the calculation of ${\rm SNR}(\bm{p}_I, \bm{w})$ in the following algorithm. Note that the only difference between $\Gamma(\cdot)$ and ${\rm SNR}(\cdot)$ is the value $\bm{p}_u$ substituted.

The overall optimization algorithm is detailed in Algorithm \ref{alg:A}, which can be summarized as follows. First, we initialize the parameters including the random UAV-IRS phase shift $\bm{w}^{[0]}$, initial UAV position $\bm{p}_I^{[0]}$, step length $k_p$ for the gradient ascent algorithm, and convergence accuracy for outer/inner loop, $\epsilon_{o}$/$\epsilon_{i}$, which should vary with changes in transmit power. Then, taking a rough estimation of UE $\hat{\bm{p}}_u$ as input, we start to optimize $\bm{p}_I$ and $\bm{w}$ alternately, where $\bm{p}_I$ is optimized using PGA-based algorithm with convergence accuracy $\epsilon_i$, and $\bm{w}$ is optimized using the proposed centroid-based method. $\bm{w}$ is considered to be fixed when optimizing $\bm{p}_I$, and vice versa. Finally, after the change of SNR is below the pre-set threshold $\epsilon_o$, the algorithm ends and outputs the optimal UAV position $\bm{p}_I^\star$ and UAV-IRS phase shift $\bm{w}^\star$.

\begin{algorithm}
\caption{Iterative joint optimization algorithm.}  
\label{alg:A}  
\begin{algorithmic}[1]

\STATE {\textbf{Input: } $\hat{\bm{p}}_u$;}
\STATE \textbf{Initialization:} $\bm{w}^{[0]}$, $\bm{p}_I^{[0]}$, $k_p$, $\epsilon_{o}$, $\epsilon_{i}$;
\STATE {$q \leftarrow 0$; $\Tilde{\bm{p}}_I^{[0]} \leftarrow \bm{p}_I^{[0]}$};
\REPEAT
    \STATE $p \leftarrow 0$;
    \REPEAT
        \STATE $p \leftarrow p + 1$;
        \STATE Update UAV position $\bm{p}_I^{[p]}$ using (\ref{updateP});
        \UNTIL{$\Gamma(\bm{p}_I^{[p]}, \bm{w}^{[q]}, \hat{\bm{p}}_u) - \Gamma(\bm{p}_I^{[p-1]}, \bm{w}^{[q]}, \hat{\bm{p}}_u) \leq \epsilon_i$};
        \STATE $q \leftarrow q + 1$;
        \STATE $\Tilde{\bm{p}}_I^{[q]} \leftarrow \bm{p}_I^{[p]}$;
        \STATE Update Phase Shift $\bm{w}^{[q]}$ using (\ref{updateW});
\UNTIL{$\Gamma(\Tilde{\bm{p}}_I^{[q]}, \bm{w}^{[q]}, \hat{\bm{p}}_u) - \Gamma(\Tilde{\bm{p}}_I^{[q-1]}, \bm{w}^{[q-1]}, \hat{\bm{p}}_u) \leq \epsilon_0$};

\STATE $\bm{p}_I^\star \leftarrow \Tilde{\bm{p}}_I^{[q]}$; $\bm{w}^\star \leftarrow \bm{w}^{[q]}$;

\STATE{\textbf{Output: } $\bm{p}_I^\star$ and $\bm{w}^\star$.}
\end{algorithmic}
\end{algorithm}  

\section{Simulation Results}

This section presents the simulation results conducted to assess the performance of the proposed optimization algorithm and its improvement to UE localization. We consider a 3D Cartesian coordinate system with its reference point located at $[0, 0, 0]^\top$. The positions of the BS and UE are $[12, 0, 2]^\top$ and $[3, 0, 1]^\top$ in meters, respectively. The initial position of UAV is $[6, 6, 3]^\top$ in meter. The combined noise power is $\sigma^2 = -125 \: {\rm dBm}$; the carrier frequency is $f_c = 28 \: {\rm GHz}$, light speed is $c = 3 \times 10^8 \: {\rm m/s}$, and the wavelength can be calculated by $\lambda = c / f_c$; the Rician factor is $\kappa = 5$; $N_B = 48$ antennas and $N_I = 48$ reflecting elements are equipped on BS and UAV-IRS, respectively, and the antenna gains for UE and BS are $G_u = G_B = 1$. The maximum moving radius of the UAV is set as $R_{max} = 3 \: {\rm m}$. Without loss of generality, we assume the UAV always moves in the plane $z = 3 \: {\rm m}$, and the UE is always located in the plane $z = 1 \: {\rm m}$. The grid search method used to solve (\ref{MLE}) searches within a $0.4 \: {\rm m} \times 0.4 \: {\rm m}$ square zone centered at the true position of UE, and uses $0.002 \: {\rm m} \times 0.002 \: {\rm m}$ grids.

For the evaluation of the proposed localization and optimization algorithms, we utilize the most commonly used \emph{root mean square error} (RMSE) and average SNR as metrics, which can be respectively expressed as:
\begin{subequations}
\begin{align}
    {\rm RMSE} &= \sqrt{\frac{1}{M} \sum_{m=1}^M \Vert \hat{\bm{p}}_{u, m} - \bm{p}_u \Vert^2}, \\
    \Bar{\rm SNR} &= \frac{1}{M} \sum_{m=1}^M {\rm SNR} (\bm{p}_{I,m}^\star, \bm{w}_m^\star),
\end{align}
\end{subequations}
where $M=300$ realizations are conducted in the simulations, $\hat{\bm{p}}_{u, m}$ represents the estimate of UE position $\bm{p}_u$ in the $m$th trail, $\bm{p}_{I,m}^\star$ and $\bm{w}_m^\star$ denote the optimal UAV position and UAV-IRS phase shift obtained in the $m$th trail, respectively.

\begin{figure}[!t]
\centering
\includegraphics[width=0.9 \linewidth]{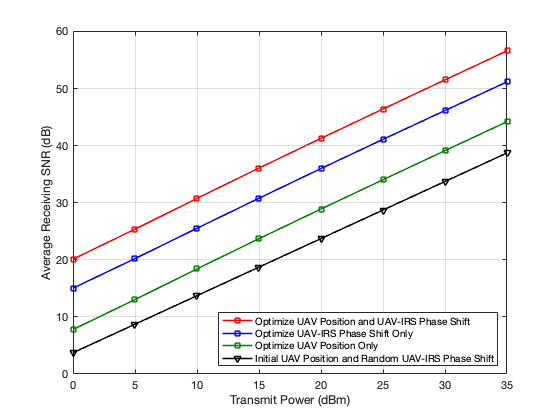}
\caption{The average receiving SNR with different UAV position and UAV-IRS phase shift design schemes.}
\label{FigSNR}
\end{figure}

Fig. \ref{FigSNR} compares the average receiving SNR under 4 different UAV position and UAV-IRS design schemes, including (a) the proposed joint optimization algorithm; (b) optimizing UAV-IRS phase shift only; (c) optimizing UAV position only and (d) using the initial UAV position and random UAV-IRS phase shift, where the optimization algorithms for schemes (b) and (c) are introduced in \ref{optimizePI} and \ref{optimizeW}, respectively. It can be observed that the average receiving SNR under all those schemes increases with transmit power. Scheme (d) results in the lowest receiving SNR because of the lack of configuration to the environment. Scheme (c) sees a slight improvement because the SNR function oscillates with the UAV position for a similar reason explained in \emph{Remark 1}, hence the PGA algorithm-based scheme can only find a local optimal value. Scheme (c) improves SNR more than scheme (b) and thus indicates that the UAV-IRS phase shift optimization algorithm outperforms the UAV position optimization algorithm. Finally, our proposed algorithm, which integrates both schemes (b) and (c), and iterates till convergence, achieves the best performance in improving the receiving SNR at BS.

\begin{figure}[!t]
\centering
\includegraphics[width=0.9 \linewidth]{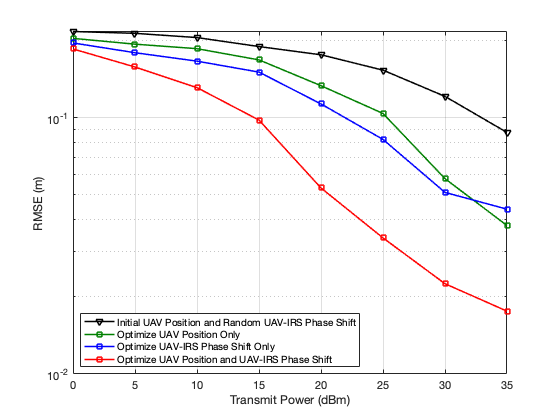}
\caption{The localization RMSE with different UAV position and UAV-IRS phase shift design schemes.}
\label{FigRMSE}
\end{figure}

Fig. \ref{FigRMSE} depicts the RMSE of UE position estimation under the same 4 schemes mentioned above. It's worth noting that the optimization process of schemes (a), (b), and (c) are based on the estimation result using scheme (d). RMSE under all schemes decreases with transmit power. The magnitude relationship of RMSE curves is almost opposite to the SNR curves, evidencing the effectiveness of maximizing receiving SNR. It's also observed that schemes (b) and (c) improve RMSE slightly, and perform similarly under the RMSE metric, while our proposed scheme (a) increases the localization accuracy significantly to nearly $10^{-2} \: {\rm m}$.

\section{Conclusion}

In this letter, we have introduced a novel UAV-IRS-assisted near-field localization system, where the UAV position can be manipulated, enhancing flexibility in configuring wireless environments and improving localization accuracy. To maximize the receiving SNR, an alternative algorithm is proposed, where the PGA algorithm is utilized to update the UAV position, and a centroid-based method is leveraged to derive closed-form expressions for UAV-RIS phase shift. Extensive simulation results indicate the effectiveness of the optimization algorithm, and the UE localization accuracy has been significantly improved to nearly $10^{-2} \: {\rm m}$ under a transmit power of $35 \: {\rm dBm}$.

\bibliographystyle{IEEEtran}
\bibliography{references}

\end{document}